\documentclass{appolb}
\usepackage{graphicx}

\begin{document}
\title{Roper resonance $N^*(1440)$ from charmonium decays %
\thanks{Submitted to the Fest Roper volume at Acta Physica Polonica B, edited by Michal Praszalowicz and
Igor Strakovsky}%
}
\author{Bing-Song Zou
\address{Department of Physics and High Energy Physics Center, Tsinghua University, Beijing 100084, China}
}
\maketitle
\begin{abstract}
The Roper resonance $N^*(1440)$ was discovered by David Roper in 1964 through sophisticated partial-wave analyses of $\pi N$ scattering data. However the first direct observation of the Roper resonance peak in the $\pi N$ invariant mass spectrum was only realized 40 years later from the charmonium decay $J/\psi\to\bar pn\pi^+ +c.c.$ at Beijing Electron-Positron Collider. The further observations of the Roper resonance production from various charmonium decays helped to reveal its multiquark nature with large $\sigma N$ component.

\end{abstract}
  
\section{Direct observation of $N^*(1440)$ from charmonium decays}

\begin{figure}
  \includegraphics[width=0.5\textwidth]{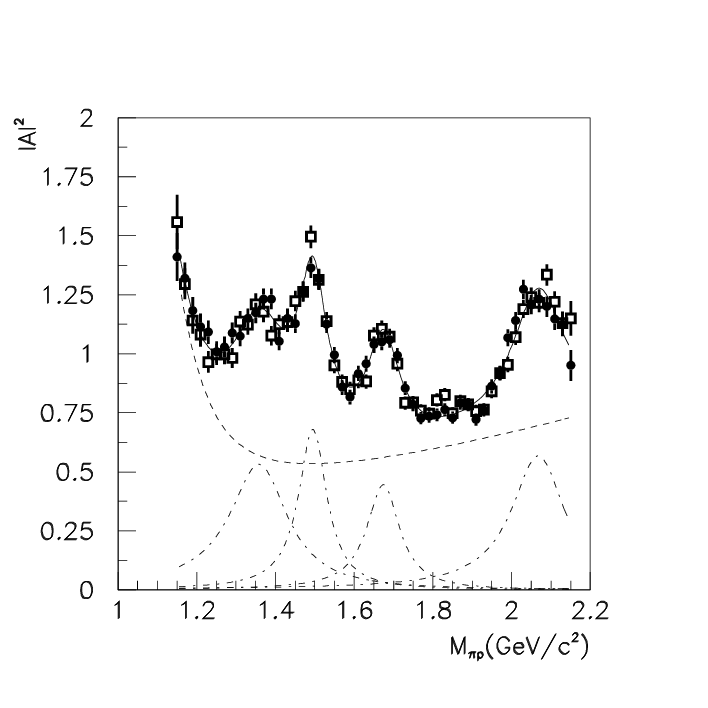}
  \includegraphics[width=0.45\textwidth]{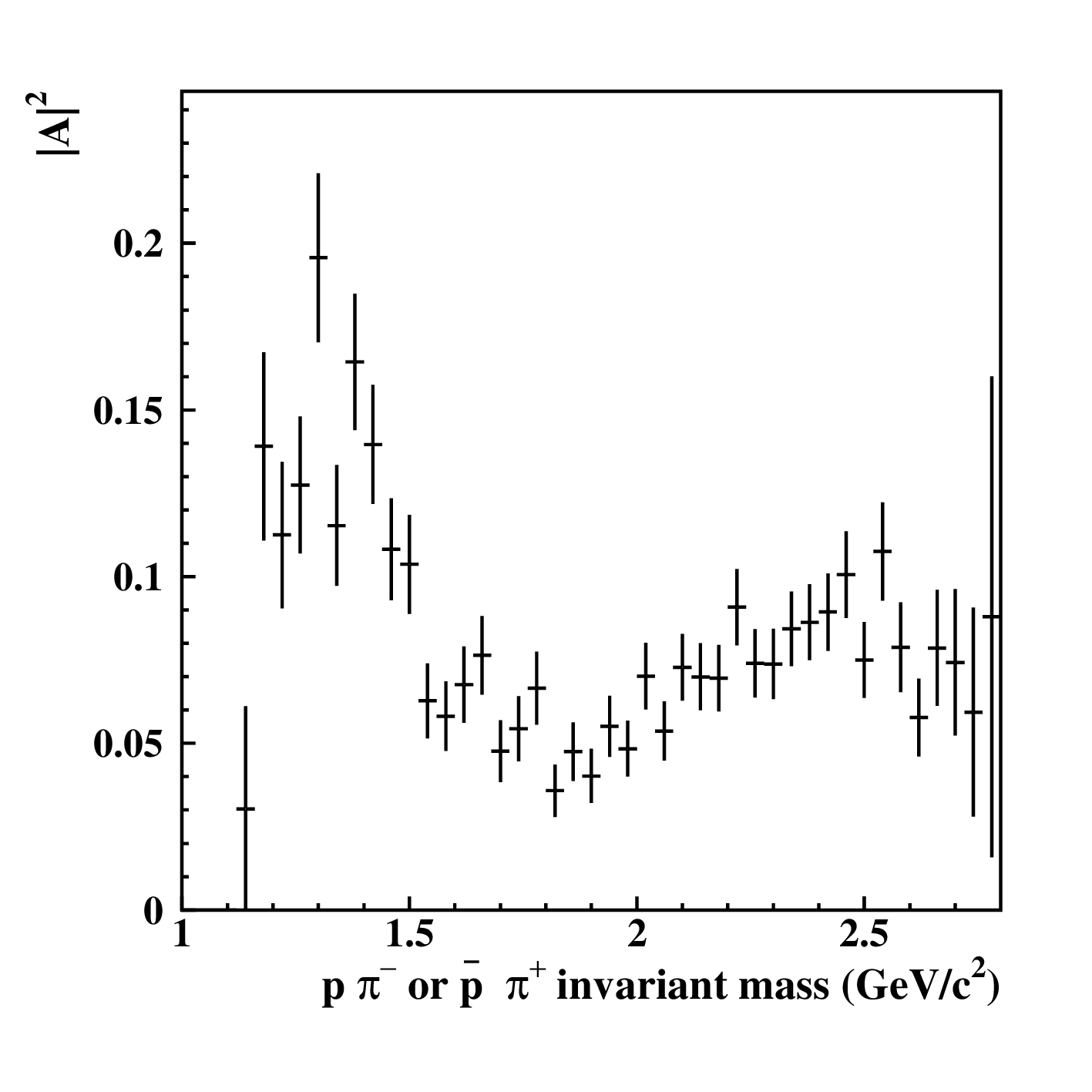}
\caption{\label{fig1} Invariant mass data corrected by MC simulated efficiency and divided by
phase space versus $p \pi^-$ (or $\bar{p} \pi^+$) invariant mass for $J/\psi \to p \bar{n}
\pi^-+c.c.$~\cite{Ablikim:2004ug} (left) and $\psi' \to p \bar{n}
\pi^-+c.c.$~\cite{Ablikim:2006aha} (right).}
\end{figure}

\begin{figure}
  \includegraphics[height=2.0in,width=4.8in]{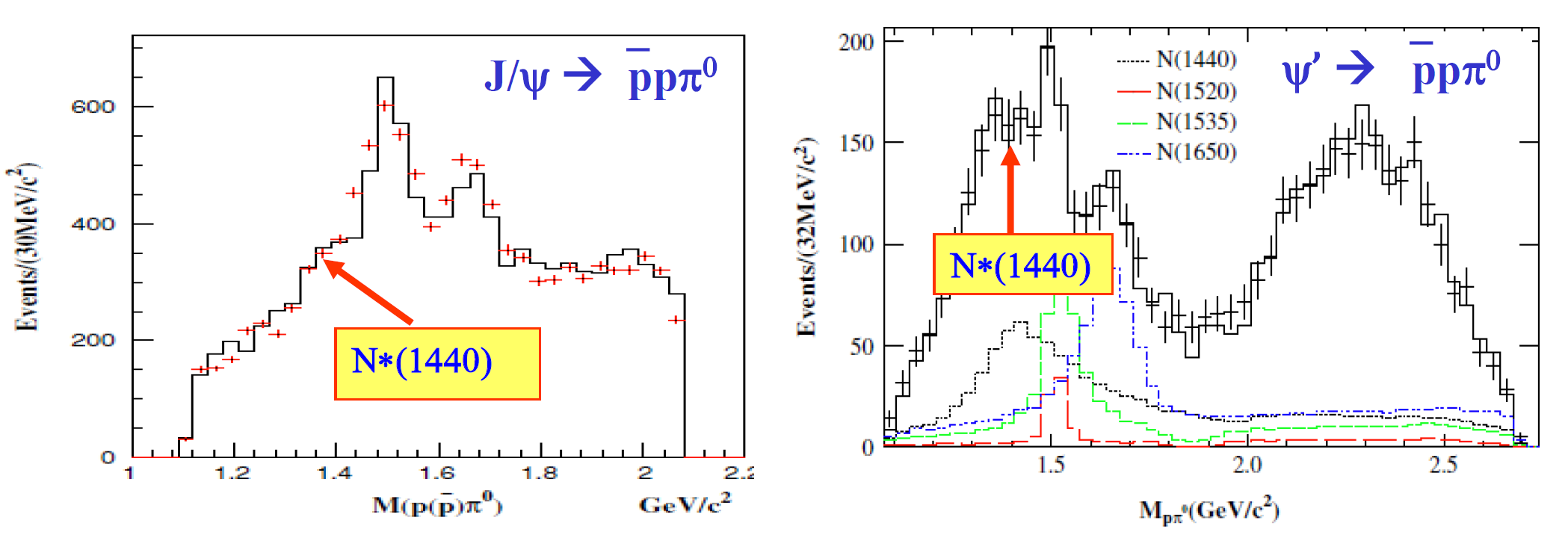}
\caption{\label{fig2} $p \pi^0$/$\bar{p} \pi^0$ invariant mass for $J/\psi \to p \bar{p}
\pi^0$~\cite{Ablikim:2009iw} (left) and $\psi' \to p \bar{p}
\pi^0$~\cite{Ablikim:2012zk} (right).}
\end{figure}

\begin{figure}
  \includegraphics[height=2.0in,width=4.8in]{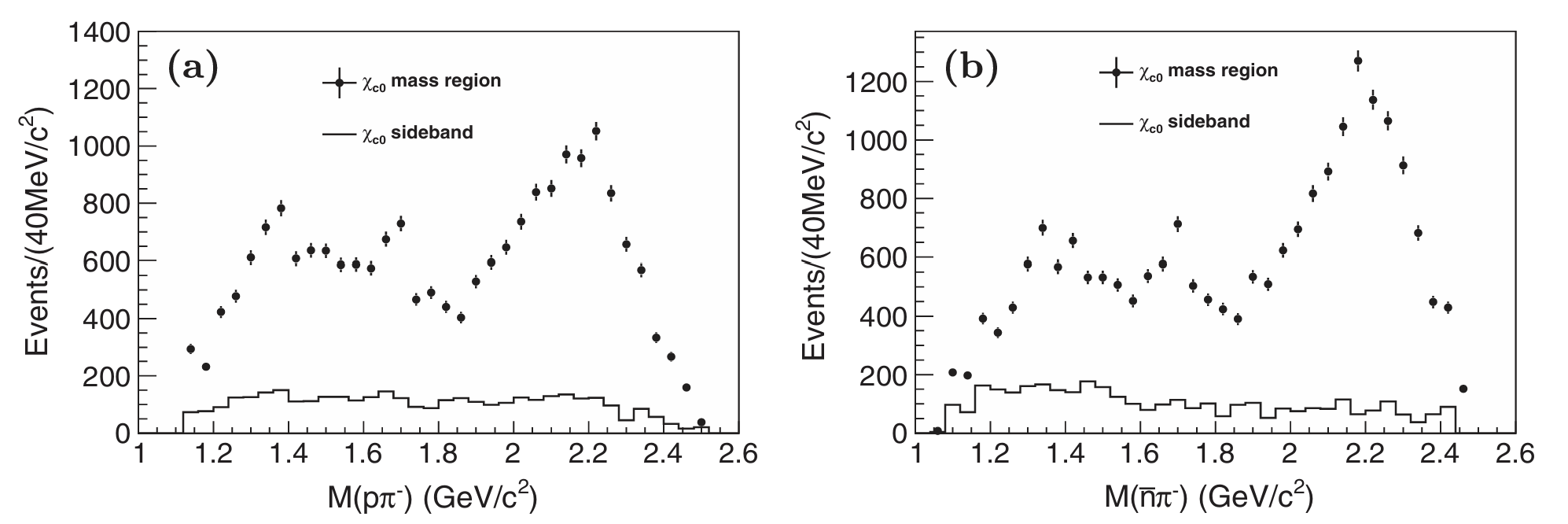}
\caption{\label{fig3}  The invariant mass distributions for (a) $p\pi^-$ and (b) $\bar n\pi^-$ of $\chi_{c0}\to p\bar n\pi^-$~\cite{BESIII:2012imn}.}
\end{figure}

Although the Roper resonance $N^*(1440)$ was discovered by David Roper long time ago~\cite{Roper:1964zza} in 1964 through sophisticated partial-wave analyses of $\pi N$ scattering data, its first direct observation in the $\pi N$ invariant mass spectrum was only realized 40 years later from the charmonium decay $J/\psi\to\bar pn\pi^+ +c.c.$ at Beijing Electron-Positron Collider~\cite{Ablikim:2004ug}. 
For the $\pi N$ and $\pi\pi N$ systems produced from $\pi N$ and $\gamma N$ reactions, they are mixture of isospin 1/2 and 3/2 with similar strengths,
and hence suffer difficulty on the isospin decomposition. In the $\pi N$ invariant mass spectrum, the $N^*(1440)$ peak is buried underneath the overwhelming $\Delta$ peak and cannot be seen directly. For the $c\bar c\to\bar NN\pi$ and $\bar NN\pi\pi$ processes, the
$\pi N$ and $\pi\pi N$ systems are expected to be dominantly isospin 1/2 due to the isospin-conserving gluon annihilation mechanism~\cite{Zou:2000wg}.
Due to the isospin filter effect, the $N^*(1440)$ peak is clearly visible in the $\pi N$ invariant mass spectra in various charmonium decays: $J/\psi\to p\bar{n}\pi^-+c.c.$~\cite{Ablikim:2004ug}, $p \bar{p}\pi^0$~\cite{Ablikim:2009iw}, and $\psi(2S)\to p \bar{n}\pi^-+c.c.$~\cite{Ablikim:2006aha}, $p \bar{p}\pi^0$~\cite{Ablikim:2012zk}, and $\chi_{c0}\to p\bar n\pi^-$~\cite{BESIII:2012imn}, as shown in Fig.\ref{fig1}, Fig.\ref{fig2} and Fig.\ref{fig3}, respectively. The fitted constant mass and width for the $N^*(1440)$ are consistent with its pole values from the PDG~\cite{ParticleDataGroup:2024cfk} obtained mainly from sophisticated partial-wave analyses of $\pi N$ scattering data of half a century. 

Later it was found that the $N^*(1440)$ peak can also be seen in the $n\pi^+$ invariant mass spectrum of $pp\to pn\pi^+$ reaction~\cite{Clement:2006ss,Ouyang:2008vg} and in the $\pi N$ invariant mass spectrum of isoscalar $NN\pi$ final state obtained from some combination of $pp\to NN\pi$ reactions~\cite{WASA-at-COSY:2017igi,Clement:2025mjg} where the $\Delta$ contribution is suppressed.
 
\section{Implication on the nature of $N^*(1440)$}

An interesting phenomena is that the $N^*(1440)$ is produced much stronger from $\psi(2S)$ than from $J/\psi$ as shown in Fig.\ref{fig1} and Fig.\ref{fig2}. This has an important implication.

There are two common features for $\psi(2S)$ and $N^*(1440)$: 1) they are supposed to be “radial” excitation of $J/\psi$ and nucleon, respectively, in the simple quenched quark model; 2) they were found experimentally to have large coupling to $\sigma J/\psi$ and $\sigma N$, respectively. In unquenched quark models, “radial” excitations like to pull out $\bar q^2q^2(0^{++})$ from sea, hence favor transition between each other. This unquenched picture not only gives a natural explanation of much enhanced $N^*(1440)$ production from $\psi(2S)$ than $J/\psi$, may also explain the long-standing $\rho\pi$ puzzle~\cite{Asner:2008nq} from $\psi(2S)$ and $J/\psi$ decays, {\sl i.e.}, $\psi(2S)$ tends to decay into $\rho(2S)\pi$ while $J/\psi$ tends to decay into $\rho\pi$. CLEO Collaboration also studied $\psi(2S)\to\bar pp\pi^0$ channel and got a similar strong $N^*(1440)$ peak~\cite{Alexander:2010vd}. There is no obvious $N^*(1440)$ produced in the $e^+e^- \to p \bar p \pi^0$ reaction with $e^+e^-$ energy around $\psi (3770)$~\cite{Ablikim:2014kxa}.

An effective field study with lattice QCD constraints~\cite{Wu:2017qve} also shows that the Roper resonance is 
best described as a resonance generated dynamically through strongly coupled meson-baryon channels with $\sigma N$ componant dominant.

It would be expected that a similar senario happens for $\Upsilon(1S,2S)\to\bar NN\pi$ with the $N^*(1440)$ produced much stronger from $\Upsilon(2S)$ than from $\Upsilon(1S)$ since the $\Upsilon(2S)$ has a large coupling to $\sigma\Upsilon(1S)$~\cite{ParticleDataGroup:2024cfk}. It is worth checking at Belle-II.

The strange partners of the $N^*(1440)$, {\sl i.e.}, $\Lambda(1600)1/2^+$ and $\Sigma(1660)1/2^+$, are clearly needed in analyzing data on $K^-p\to\pi^0\Sigma^0$~\cite{Shi:2014vha}, $K^-p\to\pi^0\Lambda$~\cite{Gao:2012zh} and $K_Lp\to\pi^+\Sigma^0$~\cite{Guo:2025mha},respectively. They are also found to have large couplings to $\Lambda\sigma$ and $\Sigma\sigma$~\cite{Sarantsev:2019xxm,ParticleDataGroup:2024cfk}, respectively. They may be further explored by the forthcoming $K_L$ beam experiment at JLab~\cite{KLF:2020gai} and Kaon beam experiments at JPARC~\cite{Aoki:2021cqa}.  

With more and more information accumulated for various "radial'' excited hadronic resonances,a consistent picture seems appearing, {\sl i.e.}, they are in fact  largely multiquark states with a $\sigma$ added to relevant ground states. 


\end{document}